\input harvmac
\input psfig
\newcount\figno
\figno=0
\def\fig#1#2#3{
\par\begingroup\parindent=0pt\leftskip=1cm\rightskip=1cm\parindent=0pt
\global\advance\figno by 1
\midinsert
\epsfxsize=#3
\centerline{\epsfbox{#2}}
\vskip 12pt
{\bf Fig. \the\figno:} #1\par
\endinsert\endgroup\par
}
\def\figlabel#1{\xdef#1{\the\figno}}
\def\encadremath#1{\vbox{\hrule\hbox{\vrule\kern8pt\vbox{\kern8pt
\hbox{$\displaystyle #1$}\kern8pt}
\kern8pt\vrule}\hrule}}
\def\underarrow#1{\vbox{\ialign{##\crcr$\hfil\displaystyle
 {#1}\hfil$\crcr\noalign{\kern1pt\nointerlineskip}$\longrightarrow$\crcr}}}
%
\overfullrule=0pt

%

\def\bar{\overline}
\def\Z{{\bf Z}}

\def\R{{\bf R}}

\font\zfont = cmss10 
\font\litfont = cmr6

\def\bigone{\hbox{1\kern -.23em {\rm l}}}
\def\ZZ{\hbox{\zfont Z\kern-.4emZ}}
\def\half{{\litfont {1 \over 2}}}

\Title{hep-th/0302194} {\vbox{\centerline{Chiral Ring Of $Sp(N)$
And $SO(N)$}
\bigskip
\centerline{ Supersymmetric Gauge Theory In Four Dimensions }}}
\smallskip
\centerline{Edward Witten}
\smallskip
\centerline{\it Institute For Advanced Study, Princeton NJ 08540 USA}

\medskip

\noindent

The chiral ring of classical supersymmetric Yang-Mills theory with
gauge group $Sp(N)$ or $SO(N)$ is computed, extending previous
work (of Cachazo, Douglas, Seiberg, and the author)
for $SU(N)$.
The result is that, as has been conjectured, the ring is generated
by the usual glueball superfield $S\sim \Tr\,W_\alpha W^\alpha$,
with the relation $S^h=0$, $h$ being the dual Coxeter number.
Though this proposition has important implications for the
behavior of the quantum theory, the statement and (for the most
part) the proofs amount to assertions about Lie groups with no
direct reference to gauge theory.

 \Date{February, 2003}

\newsec{Introduction}

\def\R{{\cal R}}
In four-dimensional supersymmetric Yang-Mills theory,  the basic
gauge invariant operator is the superspace field strength
$W_\alpha$, $\alpha=1,2$ (and its hermitian conjugate ${\bar
W}_{\dot \alpha}$). $W_\alpha$ transforms in the adjoint
representation of the gauge group, which we will take to be a
simple Lie group $G$; we denote its Lie algebra as ${\bf g}$ and
let $\Tr$ denote an invariant quadratic form on ${\bf g}$. (For
classical Lie groups, we will take $\Tr$ to be the trace in the
fundamental representation.) $W_\alpha$ is a fermionic operator of
dimension $3/2$, and is {\it chiral}, that is, it is annihilated
by the supersymmetries of one chirality: $\{\bar
Q_{\dot\alpha},W_\beta\}=0$.

Gauge-invariant polynomials in the $W_\alpha$, such as $\Tr
\,W_{\alpha_1}W_{\alpha_2}\dots W_{\alpha_s}$,  are likewise
chiral. In this paper, we consider the ``pure'' supersymmetric
gauge theory without matter multiplets.  In this theory, the
gauge-invariant polynomials in $W_\alpha$ are the only chiral
superfields of importance. However, such a polynomial is
considered trivial if it is proportional to a linear combination
of expressions $\{W_\alpha,W_\beta\}$ for any $\alpha,\beta$. The
reason for this is the identity\foot{This identity follows
directly from the superspace definition $W_\alpha=\{\bar
Q^{\dot\alpha},D_{\alpha\dot\alpha}\}$.} \eqn\uxto{\{W_\alpha,
W_\beta\}=\{\bar Q^{\dot\alpha},D_{\alpha\dot\alpha}W_\beta\},}
which implies that any gauge-invariant expression
$\sum_{\alpha\beta}X^{\alpha\beta}\{W_\alpha,W_\beta\}$ is a
descendant, that is, it can be written as $\sum_{\dot\alpha}\{\bar
Q_{\dot\alpha},Y^{\dot\alpha}\}$ for some $Y^{\dot\alpha}$, and
hence decouples from the expectation value of a product of chiral
operators.

\bigskip\noindent{\it Mathematical Description Of The Problem}

The problem of classifying modulo descendants the chiral operators
in supersymmetric gauge theory  is of mathematical as well as
physical interest. Before proceeding, let us reformulate the
problem mathematically. We introduce a $\Z_2$-graded ring $R$ that
is generated by the components of the $W_\alpha$.  Explicitly,
picking a basis $T_a$, $a=1,\dots, \dim G$ of ${\bf g}$, we write
$W_\alpha=\sum_a w^a_\alpha T_a$, and then $R$ is generated by the
(odd) variables $w_\alpha^a$.  In the ring $R$ we define an ideal
$I$ that is generated by the components of $\{W_\alpha,W_\beta\}$.
In more detail, we write $\{W_\alpha,W_\beta\}={1\over
2}\sum_{a,b}w^a_\alpha w^b_\beta [T_a,T_b]={1\over
2}\sum_{a,b,c}w^a_\alpha w^b_\beta f_{ab}^cT_c$ (where
$[T_a,T_b]=\sum_cf_{ab}^cT_c$).  Thus, $I$ is generated by the
even, nilpotent elements $\sum_{a,b}w^a_\alpha w^b_\beta
f_{ab}^c$, for all $\alpha,\beta$, and $c$.  Elements of $I$ are
descendants; the quotient ring $R/I$ is the ring of chiral
operators mod descendants.  The ideal $I$ is clearly
$G$-invariant, so $G$ acts on $R/I$.  The $G$-invariant chiral
operators mod descendants form the classical approximation to the
physical ``chiral ring'' of the theory. So in the classical
supersymmetric gauge theory, the chiral ring is $\R_{cl}=(R/I)^G$,
where $(R/I)^G$ denotes the $G$-invariant part of $R/I$, and the
subscript ``$cl$'' means ``classical'' (we recall shortly how
quantum corrections deform the picture). An element of $\R_{cl}$
can be represented by a $G$-invariant element of $R$ that is not
in $I$.

If we consider the $W_\alpha$ to be of degree one, then the ideal
$I$ is graded -- its generators being homogeneous of degree two --
and hence the classical chiral ring $\R_{cl}$ is a graded ring.
There is no non-trivial element of $\R_{cl}$ in degree one (since
there is no gauge-invariant linear function of the $W_\alpha$). In
degree two, since $W_1^2={1\over 2}\{W_1,W_1\}$ and $W_2^2={1\over
2}\{W_2,W_2\}$ are  contained in $I$, any
element of $\R_{cl}$ is a multiple of $\Tr \,W_1W_2=-\Tr\,W_2W_1$.
The degree two part of ${\cal R}_{cl}$ is thus a one-dimensional
vector space, generated by \eqn\ucvu{S=\Tr \,W_1W_2={1\over
2}\sum_\alpha\epsilon^{\alpha\beta}\Tr\,W_\alpha W_\beta.} (Here
$\epsilon^{\alpha\beta}$, $\alpha,\beta=1,2$ is the antisymmetric
tensor with $\epsilon^{12}=1$.  It is conventional to include a
factor of $-1/16\pi^2$ in the definition of $S$; this factor,
which is motivated by instanton considerations, will play no role
in the present paper and we will omit it.)

The conjecture \ref\cdsw{F. Cachazo, M. Douglas, N. Seiberg, and
E. Witten, ``Chiral Rings And Anomalies In Supersymmetric Gauge
Theory,'' hep-th/0211170, JHEP {\bf 0212:071,2002}.} that we will
be exploring in the present paper is that for any simple Lie group
$G$ with dual Coxeter number $h$, the ring $\R_{cl}$ is generated
by $S$ with the relation $S^h=0$. In \cdsw, this conjecture was
proved for $SU(N)$, and certain partial results were obtained for
other groups. For the classical Lie groups ($SO(N)$ and $Sp(N)$ as
well as $SU(N)$) it was proved that $\R_{cl}$ is generated by $S$.
(This was also proved in \ref\cere{A. Ceresole, G. Dall'Agata, R.
D'Auria, and S. Ferrara, ``Spectrum Of Type IIB Supergravity on
$AdS_5\times {\bf T}^{1,1}$: Predictions On ${\cal N}=1$ SCFT's,''
hep-th/9905226, Phys. Rev. {\bf D61} (2000) 066001.}.) For any Lie
group $G$ of rank $r$, it was proved that $S^r\not=0$ in
$\R_{cl}$. The purpose of the present paper is to prove the
conjecture for $Sp(N)$ and $SO(N)$. Most of the arguments are
similar to those in \cdsw, but for $SO(N)$ one important step in
the proof uses arguments of a quite different nature, based on
instanton calculations \ref\russians{V. A. Novikov, M. A. Shifman,
A. I. Vainshtein, M. B. Voloshin, and V. I. Zakharov,
``Supersymmetry Transformations Of Instantons,'' Nucl. Phys. {\bf
B229} (1983) 394; V. A. Novikov, M. A. Shifman, A. I. Vainshtein,
and V. I. Zakharov, ``Exact Gell-Mann-Low Functions of
Supersymmetric Yang-Mills Theories From Instanton Calculus,''
``Instanton Effects In Supersymmetric Theories,'' Nucl. Phys. {\bf
B229} (1983) 381, 407.} that were reviewed and extended in
\nref\amatietal{D. Amati, G. C. Rossi, and G. Veneziano,
``Instanton Effects In Supersymmetric Gauge Theories,'' Nucl.
Phys. {\bf B249} (1985) 1.}%
\nref\shv{M. A. Shifman and A. I. Vainshtein, ``Instantons Versus
Supersymmetry: Fifteen Years Later,'' in {\it ITEP Lectures In
Particle Physics and Field Theory}, ed M. Shifman (World
Scientific, Singapore, 1999), hep-th/9902018.} \nref\dhkm{N.
Dorey, T. J. Hollowood, V. Khoze, and M. P. Mattis, ``The Calculus
Of Many Instantons,'' hep-th/0206063, Phys. Rept.
{\bf 371} (2002) 231.}%
\refs{\amatietal -\dhkm}. For exceptional groups, a proof of the
conjecture, or even of the fact that $\R_{cl}$ is generated by
$S$, has not yet emerged.

The rings and ideals $R,I,$ and ${\cal R}_{cl}$ all admit an
action of $SL(2,{\bf C})$, under which the $W_\alpha$,
$\alpha=1,2$ transform in the two-dimensional representation.
Physically, this $SL(2,{\bf C})$ originates from an $SU(2)$
rotation symmetry of the four-dimensional gauge theory.  It will
not play an important role in the present paper.  The conjecture
about the structure of ${\cal R}_{cl}$ implies in any case that
$SL(2,{\bf C})$ acts trivially on $\R_{cl}$.

\bigskip\noindent{\it Significance For Physics}

To conclude this introduction, let us recall \cdsw\ the main
reason for the physical interest of the conjecture.  First we must
discuss the quantum deformation of the ring ${\cal R}_{cl}$. Our
definition of ${\cal R}_{cl}$ was a classical approximation to the
analogous  chiral ring ${\cal R}$ of the quantum theory. For any
$G$, the operator $S^h$, which has dimension $3h$ and carries
charge $2h$ with respect to the $U(1)_R$ symmetry of the classical
theory, has just the quantum numbers of a one-instanton
contribution to correlation functions. Hence it is possible that
the classical relation $S^h=0$ in the ring $\R_{cl}$ could be
deformed in the quantum ring ${\cal R}$  to  a relation of the
form \eqn\huto{S^h=c\Lambda^{3h}, }  with $\Lambda$ the scale
factor of the theory and some constant $c$. ($\Lambda^{3h}$ is
essentially the exponential of minus the one-instanton action and
is the standard factor that appears in all one-instanton
amplitudes.) This is the only such deformation that is possible if
the conjectured structure of $\R_{cl}$ is correct. (A
$k$-instanton amplitude has the quantum numbers of $S^{kh}$ and so
for $k>1$ cannot modify the classical relation $S^h=0$.) In fact,
explicit instanton calculations \refs{\russians - \dhkm} can be
interpreted, as we will recall in section 5, as showing that this
deformation does arise.

The quantum deformation of the classical ring $\R_{cl}$ to a
quantum ring $\R$ has a perhaps more familiar analog in two
dimensions.   In the context of two-dimensional supersymmetric
sigma models, the classical cohomology ring of (for example) ${\bf
CP}^{N-1}$, which is generated by a degree two element $x$ obeying
$x^N=0$, is naturally deformed to a quantum cohomology ring in
which the relation is $x^N=e^{-I}$.  In this case,  $I$ is the
area of a holomorphic curve of degree one (computed using a Kahler
metric on ${\bf CP}^{N-1}$ that is introduced to define the sigma
model). One difference between the four-dimensional gauge theories
and the two-dimensional sigma models is that in four dimensions
the starting point, which is the ring $\R_{cl}$ associated with a
Lie group, is less familiar mathematically than the classical
cohomology of ${\bf CP}^{N-1}$.

The quantum relation \huto\ has a striking consequence.  It
implies that in any supersymmetric vacuum, $S$ must have a nonzero
expectation value, equal to $c^{1/h}\Lambda^3$ (for one of the $h$
possible values of $c^{1/h}$). In fact, since
$\sum_{\dot\alpha}\langle \{\bar
Q_{\dot\alpha},Y^{\dot\alpha}\}\rangle=0$ for any $Y^{\dot\alpha}$
in any supersymmetric vacuum, \huto\ implies that in a
supersymmetric vacuum $\langle S^h\rangle =\langle
c\Lambda^{3h}\rangle=c\Lambda^{3h}$. But expectation values of
products of chiral operators factorize \russians, so in particular
$\langle S^h\rangle =\langle S\rangle^h$.  So we get $\langle
S\rangle^h =c\Lambda^{3h}$, whence $\langle S\rangle
=c^{1/h}\Lambda^3$, as claimed.

From this, we can deduce more.  The supersymmetric gauge theory in
four dimensions has an  anomaly-free discrete chiral symmetry
under which $S$ is rotated by an $h^{th}$ root of 1. A nonzero
expectation value of $S$ breaks this symmetry, so it follows that
in any supersymmetric vacuum, the discrete chiral symmetry is
spontaneously broken.   If a supersymmetric vacuum exists in which
$\langle S\rangle =c^{1/h}\Lambda^3$ for one choice of $c^{1/h}$,
then applying the broken symmetry gives additional vacua with the
other possible choices of $c^{1/h}$. Hence supersymmetric vacua
must come in groups of $h$, permuted by the spontaneously broken
discrete chiral symmetry.

Of course, it is believed that for any $G$, the theory has
precisely $h$ supersymmetric vacua, all with a mass gap, permuted
by the broken symmetry.  Unfortunately, no satisfactory
approximation is known for describing these vacua.

\bigskip\noindent{\it Results In The Present Paper}

In section 2 of this paper, we review the arguments given in
\cdsw. In section 3, we extend the argument for $Sp(N)$, and in
section 4, we do so for $SO(N)$.  In section 5, we briefly review
some pertinent aspects of the one-instanton calculations
\refs{\russians - \dhkm}.

\newsec{Review}

 In this section, we briefly review the known arguments.

For the classical groups $SU(N)$, $Sp(N)$,\foot{Our notation for
symplectic groups is such that $Sp(1)\cong SU(2)$.  For classical
groups, we take the symbol $\Tr$ to refer to a trace in the
fundamental representation of $SU(N)$, $Sp(N)$, or $SO(N)$ (of
dimension $N$, $2N$, and $N$, respectively).} and $SO(N)$, one can
prove directly \refs{\cere,\cdsw} that the ring ${\cal R}_{cl}$ is
generated by $S=\Tr\, W_1W_2$. In fact, for the classical groups,
any invariant polynomial in the $W_\alpha$'s is a polynomial in
the traces of words in $W_1$ and $W_2$.\foot{For $SO(2k)$, there
is an antisymmetric tensor of order $2k$, but because of
anticommutativity of $W_1$ and $W_2$, it cannot be used to make an
invariant polynomial in these variables if $k>2$.  The groups
$SO(2k)$ with $k\leq 2$ are of course not simple, so we need not
consider them.} A typical trace of such a word is
\eqn\typtr{\Tr\,W_1^{n_1}W_2^{n_2}W_1^{n_3}\dots W_2^{n_s}.} The
ideal $I$ contains $\{W_1,W_2\}$, so modulo $I$, we can take $W_1$
and $W_2$ to anticommute.  Hence the only traces to consider are
$\Tr\,W_1^{n_1}W_2^{n_2}$.  But $I$ also contains
$W_1^2={\half}\{W_1,W_1\}$, and likewise
$W_2^2={\half}\{W_2,W_2\}$. So we are reduced to generators $\Tr
\,W_1^{n_1}W_2^{n_2}$, $n_1,n_2\leq 1$. As $\Tr\,W_1=\Tr\,W_2=0$
for simple $G$, it follows that ${\cal R}_{cl}$ is generated for
$G$ a classical Lie group by $S= \Tr\, W_1W_2=-\Tr\,W_2W_1$.

The conjecture under discussion asserts more precisely that ${\cal
R}_{cl}$ is generated by $S$ with the relation $S^h=0$.  So among
other things one would like to prove that $S^{h-1}\not= 0$ in
${\cal R}_{cl}$, or equivalently, that as an element of $R$,
$S^{h-1}\notin I$.  In \cdsw, denoting the rank of $G$ as $r$, the
weaker statement that $S^r\notin I$ was proved. In fact, let ${\bf
g}={\bf t}\oplus {\bf k}$, where ${\bf t}$ is the Lie algebra of a
maximal torus in $G$, and ${\bf k}$ is its orthocomplement.  Let
$I'$ be the ideal generated by the matrix elements of the
projection of $W_\alpha $ to ${\bf k}$. Then $I\subset I'$, since
if $W_\alpha$ take values in ${\bf t}$, we have
$\{W_\alpha,W_\beta\}=0$.  To prove that $S^r\notin I$, it
suffices to show that $S^r\notin I'$.  The projection of
$W_\alpha$ to ${\bf t}$ can be written $W_\alpha=\sum_{a=1}^r
w_\alpha^a T'_a$, where the sum runs over an orthonormal basis
$T_a'$, $a=1,\dots,r$, of ${\bf t}$. The ${\bf Z}_2$ graded ring
$R/I'$ is freely generated by the odd elements $w_\alpha^a$,
$a=1,\dots,r$ (with no relations, that is, except
anticommutativity).  Moreover, $S= \sum_{a=1}^r w_1^aw_2^a$ as an
element of $R/I'$.  This formula and the absence of relations
among the $w_i^a$ make clear that $S^r\not=0$ in $R/I'$; indeed, $S^r
=r!\prod_{a=1}^rw_1^aw_2^a$.

For the Lie groups $SU(N)$ and $Sp(N)$, one has $h-1=r$. So for
these groups, the result $S^r\not=0$ in ${\cal R}_{cl}$ is
equivalent to the desired $S^{h-1}\not=0$.  For other groups,
$h-1>r$, and a direct algebraic proof that $S^{h-1}\not=0$ is not
yet known.  It is possible to use four-dimensional instantons to
prove this result; an indication of how to do so is given in
section 5.

The remaining step in \cdsw\ was to complete the proof of the
conjecture for $SU(N)$ by showing that $S^N=0$ for this group.
(For $SU(N)$, $h=N$.)  In sketching the proof, and making similar
arguments for other groups, we will make the formulas less clumsy
by writing $A$ and $B$ for $W_1$ and $W_2$.\foot{In effect, we are
here picking a basis for the space of $W_\alpha$, $\alpha=1,2$.
This will obscure the $SL(2,{\bf C})$ symmetry that was mentioned
in the introduction.} We regard $A$ and $B$ as $N\times N$
traceless matrices, and construct the following $N^{th}$ order
polynomial in $A$: \eqn\turfo{F^{i_1i_2\dots
i_N}(A)=\epsilon^{j_1j_2\dots
j_N}A^{i_1}{}_{j_1}A^{i_2}{}_{j_2}\dots A^{i_N}{}_{j_N}.} Here
$\epsilon^{j_1j_2\dots j_N}$ is the completely antisymmetric
tensor, so the right hand side of \turfo\ is antisymmetric in the
``lower'' indices of the $A$'s and hence (as the matrix elements
of $A$ anticommute) $F$ is completely symmetric in its indices
$i_1,i_2,\dots,i_N$.  As we explain in a moment, $F^{i_1i_2\dots
i_N}$ is contained in the ideal generated by matrix elements of
$A^2$. Suppose that this is known. Define the dual function of
$B$, \eqn\turgo{G_{i_1i_2\dots i_N}(B)=\epsilon_{k_1k_2\dots
k_N}B^{k_1}{}_{i_1}B^{k_2}{}_{i_2}\dots B^{k_N}{}_{i_N}.} It is
likewise contained in the ideal generated by $B^2$.  Since $F$ and
$G$ are both contained in the ideal $I$, so is
\eqn\jundo{\eqalign{F(A)\cdot G(B)=&F^{i_1i_2\dots
i_N}(A)G_{i_1i_2\dots i_N}(B)\cr =&\epsilon^{j_1j_2\dots
j_N}A^{i_1}{}_{j_1}A^{i_2}{}_{j_2}\dots
A^{i_N}{}_{j_N}\epsilon_{k_1k_2\dots
k_N}B^{k_1}{}_{i_1}B^{k_2}{}_{i_2}\dots B^{k_N}{}_{i_N}.\cr}} But
a direct evaluation of the right hand side of \jundo\ can be made
using the identity \eqn\fundo{\epsilon^{j_1j_2\dots
j_N}\epsilon_{k_1k_2\dots
k_N}=\delta^{j_1}_{k_1}\delta^{j_2}_{k_2}\dots
\delta^{j_N}_{k_N}\pm {\rm permutations~of~}k_1,k_2,\dots k_N.}
When this is done, all indices of $A$'s become contracted with
indices of $B$'s, implying that  $F(A)\cdot G(B)$ is a sum of
terms $\Tr (AB)^{r_1}\Tr(AB)^{r_2}\dots\Tr (AB)^{r_m}$, with
various $r_i$.  The coefficient of $S^N=(\Tr\,AB)^N$ is nonzero --
it is 1, coming from the trivial permutation in \fundo. The other
terms with some $r_i>1$ are contained in $I$, as we have seen in
proving that $\R_{cl}$ is generated by $S$. Hence $S^N\in I$.

So it remains only to show that $F(A)^{i_1i_2\dots i_N}$ is in the
ideal generated by $A^2$.  Without loss of generality, since this
tensor is symmetric in the indices $i_k$, we can set these to a
common value, say $N$.  We will show that
\eqn\inco{\epsilon^{j_1j_2\dots j_N}A^N{}_{j_1}A^N{}_{j_2}\dots
A^N{}_{j_N}} is a nonzero multiple of
\eqn\binco{\epsilon^{Nj_1j_2\dots
j_{N-1}}(A^2)^N{}_{j_1}A^N{}_{j_2}A^N{}_{j_3}\dots
A^N{}_{j_{N-1}},} which is certainly proportional to $A^2$. We can
write \binco\ more explicitly as
\eqn\dinco{\sum_{x=1}^N\epsilon^{Nj_1j_2\dots j_{N-1}}A^N{}_x
A^x{}_{j_1}A^N{}_{j_2}A^N{}_{j_3}\dots A^N{}_{j_{N-1}}.} The
expression \eqn\tinco{A^N{}_xA^N{}_{j_2}A^N{}_{j_3}\dots
A^N{}_{j_{N_1}},} being antisymmetric in $x,j_2,\dots,j_{N-1}$, is
a nonzero multiple of \eqn\flinco{\epsilon_{xj_2j_3\dots
j_{N-1}r}\epsilon^{rs_1s_2\dots
s_{N-1}}A^N{}_{s_1}A^N{}_{s_2}\dots A^N{}_{s_{N-1}}.} Now
substitute this expression in \dinco, and then use \fundo\ to
write the product $\epsilon^{Nj_1j_2\dots
j_{N-1}}\epsilon_{xj_2j_3\dots j_{N-1}r}$ as a multiple of
$\delta^N_x\delta^{j_1}_r-\delta^N_r\delta^{j_1}_x$.   We learn
that \dinco\ is a nonzero multiple of
\eqn\polyo{\left(\delta^N_x\delta^{j_1}_r-\delta^N_r\delta^{j_1}_x\right)
A^x{}_{j_1}\epsilon^{rs_1s_2\dots
s_{N-1}}A^N{}_{s_1}A^N{}_{s_2}\dots A^N{}_{s_{N-1}}.} The
$\delta^{j_1}_x$ terms give a multiple of $\Tr\,A$, which vanishes
for $A$ in the Lie algebra of $SU(N)$, and the
$\delta^N_x\delta^r_{j_1}$ term gives \inco, as promised.

\newsec{Proof For $Sp(N)$}

In this section, we prove the conjecture for $G=Sp(n)$.  For this
group, $h=N+1$ and $r=h-1=N$.  Since we have in section 2
explained why $\R_{cl}$ is generated by $S$ for $Sp(N)$, and why
$S^N=S^r\not=0$, we need only prove that $S^{N+1}=0$.

We recall that a generator $A$ of $Sp(N)$ can be represented as a
$2N\times 2N$ symmetric tensor $A_{ij}$.  Indices are raised and
lowered using the invariant antisymmetric tensor $\gamma_{jk}$ of
$Sp(N)$, and its inverse $\gamma^{kr}$:
$A^i{}_j=\gamma^{ik}A_{kj}$, $A_{ij}=\gamma_{ik}A^k{}_j$, with
$\gamma_{ik}\gamma^{kj}=\delta^j_i$.  The definition of $S$ is
$S=\Tr\,AB=A_{ij}B_{kl}\gamma^{jk}\gamma^{li}$. To think of $A$ as a matrix that
can be multiplied, one should raise an index and use $A^i{}_j=\gamma^{ik}A_{kj}$.
The ideal $I$ is
generated by the matrix elements of $A^2$, or explicitly by the
quantities \eqn\ormi{\sum_{kl}A_{ik}A_{jl}\gamma^{kl},} as well as
similar expressions with one or both $A$'s replaced by $B$.

The antisymmetric tensor $\gamma^{ij}$ is nondegenerate and has a nonzero
Pfaffian.  This implies that the antisymmetric tensor
$\epsilon^{i_1i_2\dots i_{2N}}$ can be written in terms of
$\gamma$: \eqn\jugo{\epsilon^{i_1i_2\dots
i_{2N}}=\gamma^{i_1i_2}\gamma^{i_3i_4}\dots
\gamma^{i_{2N-1}i_{2N}} \pm {\rm permutations}.}

The strategy of the proof will be the same as for $SU(N)$.  We
will construct a polynomial $F(A)$ which is contained in the ideal
generated by $A^2$, and which when contracted with the analogous
polynomial in $B$ is equal to $S^{N+1}$ modulo $I$. We simply set
\eqn\truro{F_{i_1i_2\dots i_{N+1}}^{k_1k_2\dots
k_{N-1}}(A)=\epsilon^{k_1k_2\dots k_{N-1}j_1j_2\dots
j_{N+1}}A_{i_1j_1}A_{i_2j_2}\dots A_{i_{N+1}j_{N+1}}.} $F$ is
antisymmetric in the $k$'s and symmetric in the $i$'s.  To show
that $F(A)$ is contained in the ideal $I$, we use the identity
\jugo\ to express the tensor $\epsilon$ as a sum of products of
$N$ $\gamma$'s.  The $\gamma$'s have a total of $2N$ indices,
$N+1$ of which are $j_1,j_2,\dots,j_{N+1}$ and are contracted with
$A$'s. $N+1$ exceeds the number of $\gamma$'s, so
 in each term of the sum, at least one $\gamma$ has two
indices $j_m,j_n$ that are contracted with $A$'s.  Since a
$\gamma$ cannot be contracted twice with the same $A$
($\gamma^{ij}A_{ij}=0$ as $A$ is symmetric), the $\gamma$ in
question is contracted once each with two different $A$'s, giving
$A_{i_mj_m}A_{i_nj_n}\gamma^{j_mj_n}$, for some values of the
indices; this is a generator of $I$. So $F(A)\in I$.

After defining $F_{i_1i_2\dots i_{N+1}}^{k_1k_2\dots k_{N-1}}(B)$
by the same formula, we now want to evaluate \eqn\yco{F(A)\cdot
F(B)=F^{i_1i_2\dots i_{N+1}}_{k_1k_2\dots
k_{N-1}}(A)F_{i_1i_2\dots i_{N+1}}^{k_1k_2\dots k_{N-1}}(B).} (The
indices of $F(A)$ have been raised and lowered with $\gamma$'s to
make this contraction.)  Clearly, $F(A)\cdot F(B)$ is contained in
$I$, since $F(A) $ and $F(B)$ are.  However, we can also evaluate
$F(A)\cdot F(B)$ by working directly from the definition:
\eqn\ucu{\eqalign{F(A)\cdot F(B)=&\epsilon_{k_1k_2\dots
k_{N-1}j_1j_2\dots j_{N+1}}A^{i_1j_1}A^{i_2j_2}\dots
A^{i_{N+1}j_{N+1}}\cr &\cdot \epsilon^{k_1k_2\dots
k_{N-1}m_1m_2\dots m_{N+1}}B_{i_1m_1}B_{i_2m_2}\dots
B_{i_{N+1}m_{N+1}}.\cr}} Now upon using the identity \fundo\ to
evaluate $\epsilon^{k_1k_2\dots k_{N-1}m_1m_2\dots
m_{N+1}}\epsilon_{k_1k_2\dots k_{N-1}j_1j_2\dots j_{N+1}}$, all
indices of $A$'s are contracted with indices of $B$'s, and we get
as in section 2 a sum of terms each of which is of the form
$\Tr\,(AB)^{r_1}\,\Tr\,(AB)^{r_2}\dots\Tr\,(AB)^{r_m}$ for some
$r_i$.  The coefficient of $S^{N+1}=(\Tr\,AB)^{N+1}$ is nonzero,
and the other terms with some $r_i>1$ are again all contained in
the ideal $I$.  So we have shown that $S^{N+1}$ is contained in
$I$, completing the proof of the conjecture for the symplectic
group.

\newsec{Proof For $SO(N)$}

For $SO(N)$, the dual Coxeter number is $h=N-2$.  The proof that
$S^{N-2}\in I$ will be similar to what we have already seen,
though slightly more elaborate.  A novelty for $SO(N)$ is that
$h>r+1$ in this case, so the argument using reduction to a maximal
torus (which only shows that $S^r\notin I$) does not suffice to
show that $S^{h-1}\notin I$. The only proof of this that I know of
uses facts about four-dimensional instantons and is deferred to
section 5.

An element of the Lie algebra of $SO(N)$ is an antisymmetric
$N\times N$ matrix $A_{ij}$; indices are raised and lowered and contracted using the
invariant metric  $\delta_{ij}$ and its inverse $\delta^{ij}$.
Since indices can be raised and lowered in a unique way without introducing
any minus signs, we make no distinction between upper and
lower indices. The ideal $I$ is generated by
\eqn\ugu{(A^2)_{ij}=\sum_kA_{ik}A_{kj}=-\sum_kA_{ik}A_{jk},} and
analogous expressions with one or both $A$'s replaced by $B$.
Apart from $\delta_{ij}$, the only independent invariant tensor is
the antisymmetric tensor $\epsilon_{i_1i_2\dots i_N}$.

Since the proof that $S^{N-2}\in I$ will be slightly elaborate, we
first consider the case of $SO(5)$ (which is isomorphic to $Sp(2)$
so that we could borrow the result of the last section, though the
argument will not be expressed in such terms).  To prove that
$S^3=0$ for $SO(5)$, we will construct a cubic polynomial $F(A)$,
which is contained in $I$ and when contracted with the analogous
cubic polynomial in $B$ is equal to $S^3 $ mod $I$. This will show
that $S^3\in I$ for $SO(5)$.  Then we will generalize the
construction to $SO(N)$ with $N>5$.  (Since $SO(3)$ is equivalent
to $SU(2)$ and $SO(4)$ to $SU(2)\times SU(2)$, we need not
consider those cases.)

To construct $F(A)$, we will begin with a product of three $A$'s,
say $A_{rr'}A_{ss'}A_{tt'}$, and a product of two antisymmetric
tensors,  $\epsilon_{i_1i_2\dots i_5}\epsilon_{j_1j_2\dots j_5}$.
Then we will contract all six indices of the $A$'s with some of
the ten indices carried by the antisymmetric tensors.  There is
essentially only one way to do this.  We cannot take two $A$'s and
contract all four of their indices with the same antisymmetric
tensor, since
\eqn\ogu{A_{i_1i_2}A_{i_3i_4}\epsilon_{i_1i_2i_3i_4i_5}=0} by
anticommutativity.  Likewise, we cannot have two $A$'s each with
one index contracted with each of the antisymmetric tensors, since
again
\eqn\bogu{A_{i_1j_1}A_{i_2j_2}\epsilon_{i_1i_2i_3i_4i_5}\epsilon_{j_1j_2j_3j_4j_5}=0}
by anticommutativity. So the only nonzero expression that we can
make by contracting all six indices of the three $A$'s with six of
the ten indices of the antisymmetric tensors is
\eqn\hogu{F_{i_1i_2j_1j_2}(A)=A_{i_3i_4}A_{j_3j_4}A_{i_5j_5}
\epsilon_{i_1i_2i_3i_4i_5}\epsilon_{j_1j_2j_3j_4j_5},} in which
one $A$ is contracted twice with the first antisymmetric tensor,
one is contracted twice with the second, and one is contracted
once with each.

If we insert in the definition of $F$ the identity
$\epsilon_{i_1i_2i_3i_4i_5}\epsilon_{j_1j_2j_3j_4j_5}=\delta_{i_1j_1}\delta_{i_2j_2}\dots
\delta_{i_5j_5}\pm {\rm permutations}$, we get a sum of many terms
each proportional to a product of five metric tensors
$\delta_{i_mj_n}$.  The five metrics have a total of ten indices,
six of which are contracted with indices of the three $A$'s.  The crucial fact
is that six exceeds five, so one metric has both indices contracted with $A$'s.  One
cannot contract a metric tensor twice with the same $A$
($\delta_{ij}A_{ij}=0$, as $A$ is antisymmetric).  So inevitably,
in each term, one of the $\delta$'s is contracted with two
different $A$'s, giving an expression $A_{mn}\delta_{np}A_{pr}=(A^2)_{mr}$
that is a generator of the ideal $I$.  So $F(A)$ is contained in $I$.

On the other hand, consider \eqn\tufno{F(A)\cdot
F(B)=A_{i_3i_4}A_{j_3j_4}A_{i_5j_5}
\epsilon_{i_1i_2i_3i_4i_5}\epsilon_{j_1j_2j_3j_4j_5}
B_{k_3k_4}B_{t_3t_4}B_{k_5t_5}
\epsilon_{i_1i_2k_3k_4k_5}\epsilon_{j_1j_2t_3t_4t_5}.} We can
evaluate this by using \fundo\ to express the products
$\epsilon_{i_1i_2i_3i_4i_5} \epsilon_{i_1i_2k_3k_4k_5}$ and $
\epsilon_{j_1j_2j_3j_4j_5}\epsilon_{j_1j_2t_3t_4t_5}$ in terms of
products of $\delta$'s.  When we do this,  the indices of $A$'s
and $B$'s are contracted, and we get a sum of terms, each of which
is a product of traces of words in $A$ and $B$.  The sum includes
a positive multiple of $S^3$, and additional terms that are
contained in $I$ because the trace of any word with more than two
letters is in $I$.  This proves that $S^3\in I$ for $SO(5)$.

To prove in a similar fashion that $S^{N-2}\in I$ for $SO(N)$, we
should start with $N-2$ factors of $A$ and contract some of their
indices with a product of two $\epsilon$'s to define a polynomial
$F(A)$.  To prove along the above lines that $F(A)\in I$, we need
to have at least $N+1$ indices of the product of $\epsilon$'s
contracted with $A$'s.  Let us verify that this is just possible.
As we have seen above, three $A$'s can be contracted twice each
with the product of $\epsilon$'s, giving a total of 6
contractions. The remaining $N-5$ $A$'s can each have only one
index contracted with the product of $\epsilon$'s, since two
contractions will give a vanishing result by virtue of \ogu\ or
\bogu\ (which have obvious analogs for $N>5$).  The total number
of contractions will hence be $6+(N-5)=N+1$, exactly what we need.
It does not matter with which antisymmetric tensor the last $N-5$
$A$'s are contracted.  So we define
\eqn\wedef{\eqalign{F(A)&_{i_1i_2j_1j_2\dots j_{N-3}s_1s_2\dots
s_{N-5}}\cr&=\epsilon_{i_1i_2t_1t_2\dots
t_{N-2}}\epsilon_{j_1j_2\dots
j_{N-3}k_1k_2k_3}A_{t_1t_2}A_{k_1k_2}A_{t_3k_3}A_{t_4s_1}A_{t_5s_2}\dots
A_{t_{N-2}s_{N-5}}.\cr}} $F(A)$ is contained in $I$ for the
familiar reason: upon using \fundo\ to replace the product of
antisymmetric tensors with a sum of products of $\delta$'s, we get
a sum of terms in each of which some $\delta$ is contracted with
two $A$'s, giving a generator of $I$.

Hence the quantity \eqn\guyy{F(A)\cdot
F(B)=F(A)_{i_1i_2j_1j_2\dots j_{N-3}s_1s_2\dots
s_{N-5}}F(B)_{i_1i_2j_1j_2\dots j_{N-3}s_1s_2\dots s_{N-5}}} is
contained in $I$.  On the other hand, explicitly
\eqn\nuyy{\eqalign{F(A)\cdot F(B)=\,\,&\epsilon_{i_1i_2t_1t_2\dots
t_{N-2}}\epsilon_{j_1j_2\dots
j_{N-3}k_1k_2k_3}A_{t_1t_2}A_{k_1k_2}A_{t_3k_3}A_{t_4s_1}A_{t_5s_2}\dots
A_{t_{N-2}s_{N-5}}\cr &\epsilon_{i_1i_2u_1u_2\dots
u_{N-2}}\epsilon_{j_1j_2\dots
j_{N-3}n_1n_2n_3}B_{u_1u_2}B_{n_1n_2}B_{u_3n_3}B_{u_4s_1}B_{u_5s_2}\dots
B_{u_{N-2}s_{N-5}}.\cr}} Using \fundo\ to replace
$\epsilon_{i_1i_2t_1t_2\dots t_{N-2}}\epsilon_{i_1i_2u_1u_2\dots
u_{N-2}}$ and likewise
    $\epsilon_{j_1j_2\dots j_{N-3}k_1k_2k_3}\epsilon_{j_1j_2\dots
j_{N-3}n_1n_2n_3}$ with sums of products of $\delta$'s, we learn
in the familiar fashion that $F(A)\cdot F(B)$ is equal to
$S^{N-2}$ plus a sum of terms (proportional to traces of longer
words in $A$ and $B$) that are contained in $I$.  Combining these
results, we deduce that $S^{N-2}\in I$.

\newsec{Implications Of Instanton Calculations}

For any simple Lie group $G$, a one-instanton solution on ${\bf
R}^4$ is obtained by picking a minimal $SU(2)$ subgroup of $G$,
and embedding in $G$ the one-instanton solution of $SU(2)$.  Under
such a minimal $SU(2)$, the Lie algebra ${\bf g}$ of $G$
decomposes as the adjoint representation of $SU(2)$ plus a certain
number of pairs of copies of the spin one-half representation, as
well as $SU(2)$ singlets. Because the same $SU(2)$ representations
arise for any $G$, the relevant properties of the one-instanton
computation are largely independent of $G$.  A computation for all
simple Lie groups was performed in \ref\dhk{N. M. Davies, T. J.
Hollowood, and V. V. Khoze, ``Monopoles, Affine Algebras, and The
Gluino Condensate,'' hep-th/0006011.}.

For instanton number one, the instanton moduli are the position
and size and ``$SU(2)$ orientation'' of the instanton and the
choice of minimal embedding of $SU(2)$ in $G$.

In the field of the instanton, the gluino field (which is a fermi
field with values in the adjoint representation of $G$) has $2h$
zero modes, all of one chirality.  This is the right number to
give an expectation value to an operator with the quantum numbers
of a product of $h$ copies of $S$.  In \russians, general
properties of chiral operators were used to show that in a
supersymmetric vacuum the expectation value of a product of chiral
operators such as  $\langle S(x_1)S(x_2)\dots S(x_h)\rangle$ is
independent of the choice of points $x_i\in{\bf R}^4$ as long as
the $x_i$ are distinct, ensuring there are no ambiguities in
defining the operator products.  Moreover, a one-instanton
computation was performed on ${\bf R}^4$, with the result
\eqn\pinno{\langle S(x_1)S(x_2)\dots S(x_h)\rangle_{1~{\rm
inst}}=c_0\Lambda^{3h}} for some constant $c_0$. The computation
is made by evaluating $S(x_i)$ as bilinear expressions in the
fermion zero modes (corrections to this vanish by holomorphy) and
then integrating over instanton moduli space. The subscript
``$1~{\rm inst}$'' in \pinno\ refers to the fact that we are
recording here the result of a one-instanton computation, which
may or may not give the exact quantum answer.

Our main goal in the present section is to argue from properties
of the one-instanton moduli space that $S^{h-1}\notin I$.  For
this, we do not need to know whether the one-instanton computation
gives the exact quantum answer or not; in fact, we do not even
need to know if the quantum theory really exists.  The argument we
will give could be formulated as a conventional mathematical proof
that $S^{h-1}\notin I$, using properties of instanton moduli
space.

We will also sketch how instantons are used to deduce the quantum
anomaly that makes $S^h$ a non-zero multiple of the identity  in
the quantum chiral ring (rather than vanishing, as it does
classically). For this, one does need to know something about the
quantum theory, so after arguing that $S^{h-1}\notin I$, we will
recall some issues concerning the relation of the instanton
computation to the quantum theory.

It is possible to take $h-1$ of the $x_i$ to coincide without
running into any difficulty or ambiguity and in particular without
running into a singular contribution from small instantons.  (See
eqn. (7.17) of \dhkm, where this choice is made.) So \eqn\truto{
\langle S(x)S^{h-1}(y)\rangle_{1~{\rm inst}}=c_0\Lambda^{3h}.}
From this we can deduce the desired result that $S^{h-1}\notin I$.
Indeed, because of the $\bar Q$-invariance of the one-instanton
computation, a formula $S^{h-1}=\{\bar
Q_{\dot\alpha},Y^{\dot\alpha}\}$ would imply the vanishing of
\truto\ (it would lead to a representation of \truto\ as the
integral of a total derivative over instanton moduli space). Since
it does not vanish, $S^{h-1}\notin I$.

This completes what we have to say about nonvanishing of $S^{h-1}$
in the classical theory. Now let us discuss how the quantum
anomaly in $S^h$ comes about. If we simply set $x=y$ in \truto, we
find that the function that must be integrated over instanton
moduli space is identically zero.  The reason for this is that at
each point $y\in {\bf R}^4$ and for any given one-instanton
solution, two of the fermion zero modes vanish.  (They are a
suitable linear combination, depending on $y$ and on the position
of the instanton, of the zero modes generated by global
supersymmetries and superconformal transformations.)  Hence, when
$S^h(y)$ is evaluated using the fermion zero modes, one gets
identically zero before doing any integral over instanton moduli
space.  This is compatible with (but stronger than) the kind of
behavior of $S^h$ that one would expect from the classical result
$S^h\in I$ (this result would make us expect a perhaps not
identically zero total derivative on moduli space).

In the quantum theory, however, we should be careful in defining
an operator product such as $S^h$.  This is conveniently done by
point-splitting, taking a product such $S(x_1)S(x_2)\dots S(x_h)$
and, after performing the computation, taking the limit as the
$x_i$ coincide.  In the present case, there is no problem in
setting $h-1$ of the $x_i$ equal (since no singular
small-instanton contributions appear in \truto\ as long as $x\not=
y$).  But we should be careful to define $S^h(y)$ as $\lim_{x\to
y} S(x)S^{h-1}(y)$.  When we do this, clearly we get \eqn\thyr{
\langle S^h(y)\rangle_{1~{\rm inst}}=c_0\Lambda^{3h}.}

If we assume that the one-instanton amplitude coincides with the
exact quantum answer, we would deduce from this that the classical
ring relation $S^h=0$ is deformed quantum mechanically to
$S^h=c_0\Lambda^{3h}$.  However, it is believed that the exact
quantum answer is actually \eqn\gry{\langle
S^h(y)\rangle=c\Lambda^{3h},} with a {\rm different} constant $c$,
so that the quantum ring relation is really $S^h=c\Lambda^{3h}$.
The  discrepancy between the one-instanton computation of the
anomaly coefficient and the exact result is still somewhat
surprising; for a detailed analysis and references, see section 7
of \dhkm.  One simple statement \ref\hk{N. M. Davies, T. J.
Hollowood, V. V. Khoze, and M. P. Mattis, ``Gluino Condensate And
Magnetic Monopoles In Supersymmetric Gluodynamics,'' Nucl. Phys.
{\bf B559} (1999) 123, hep-th/9905015.} is that if the
one-instanton computation is done on ${\bf R}^3\times {\bf S}^1$
instead of ${\bf R}^4$, with an arbitrary radius  for the ${\bf
S}^1$, one gets the result \gry, with what is believed to be the
correct coefficient $c$, independent of the radius (as long as the
radius is finite). Since the statement $S^h=c\Lambda^{3h}+\{\bar
Q_{\dot\alpha},Y^{\dot\alpha}\}$ is an operator statement,
independent of any particular choice of state, the coefficient $c$
can be computed, in principle, on any chosen four-manifold and
with any chosen boundary conditions. Compactification on ${\bf
S}^1$ with small radius and a non-trivial Wilson loop expectation
value at infinity gives a suitable  framework for a reliable
computation of the anomaly coefficient in a weakly coupled
context.  (The proof that $S^{h-1}\notin I$ could also have been
carried out in just the same way after compactification on ${\bf
S}^1$.)  The direct computation on ${\bf R}^4$ is presumably
affected by some infrared divergences in the relation between the
perturbative vacuum in which the computation is done and the true
quantum vacuum.

\bigskip\noindent{Note Added In hep-th version 3}

In a paper \ref\kac{P. Etinghof and V. Kac, ``On The
Cachazo-Douglas-Seiberg-Witten Conjecture For Simple Lie
Algebras,'' math.QA/0305175.} that appeared some months after the
original hep-th version of the present one, Etinghof and Kac have
verified the conjectured structure of the classical chiral ring
for the exceptional Lie group $G_2$.

\bigskip
This work was supported in part by NSF Grant PHY-0070928.  I would
like to acknowledge the hospitality of the particle theory group
at Caltech, where much of this work was done, and to thank T.
Hollowood, V. Kac, V. Khoze, and  N. Seiberg for discussions.

 \listrefs
\end